\documentclass[conference]{IEEEtran}
\IEEEoverridecommandlockouts
\usepackage{cite}
\usepackage{amsmath,amssymb,amsfonts}
\usepackage{algorithmic}
\usepackage{graphicx}
\usepackage{textcomp}
\usepackage{xcolor}
\def\BibTeX{{\rm B\kern-.05em{\sc i\kern-.025em b}\kern-.08em
    T\kern-.1667em\lower.7ex\hbox{E}\kern-.125emX}}
\begin{document}

\title{Labeling Questions Inside Issue Trackers\\
% {\footnotesize \textsuperscript{*}Note: Sub-titles are not captured in Xplore and should not be used}
}

\author{\IEEEauthorblockN{Aidin Rasti}
\IEEEauthorblockA{\textit{Electrical Engineering and Computer Science} \\
\textit{University of Ottawa}\\
Ottawa, Canada \\
arast040@uottawa.ca}
}

\maketitle

% \begin{abstract}
% This document is a model and instructions for \LaTeX.
% This and the IEEEtran.cls file define the components of your paper [title, text, heads, etc.]. *CRITICAL: Do Not Use Symbols, Special Characters, Footnotes, 
% or Math in Paper Title or Abstract.
% \end{abstract}

\begin{abstract}
  One of the issues faced by the maintainers of popular open source software is the triage 
  of newly reported issues. Many of the issues submitted to issue trackers are questions.
  Many people ask questions on issue trackers about their problem instead of using a proper QA 
  website like StackOverflow. This may seem insignificant but for many of the big projects 
  with thousands of users, this leads to spamming of the issue tracker. Reading and labeling
  these unrelated issues manually is a serious time consuming task and these unrelated questions 
  add to the burden. In fact, most often maintainers demand to not submit questions in the 
  issue tracker.

  To address this problem, first, we leveraged dozens of patterns
  to clean text of issues, we removed noises like logs, stack traces, environment variables, 
  error messages, etc.
  Second, we have implemented a classification-based approach to automatically label unrelated questions. 
  Empirical evaluations on a dataset of more than 102,000 records show that our approach 
  can label questions with an accuracy of over 81\%. 
\end{abstract}  


\section{Introduction}
With the prevalence of open source software, authors of projects kindly carry on 
the responsibility of supporting users and providing documentation. The relation 
between users and authors is a two-way relation, developers rely on feedback and reports from 
users to improve their software. These feedback reports come in the form of bugs, questions, 
features, suggestions, and enhancements, or what we technically call "issue"s. Most developers
prefer to use the issue tracking software just as a means for developing the software, it is preferred 
that questions related to the workings or documentation of the software be directed to forums or QA
websites. And by questions we mean everything that is, asking about how to fix an error/problem, 
asking about features and documentation, asking for help, etc.

These unrelated questions only add to the clutter of issue trackers, especially in bigger projects.
Developers have to put a tremendous amount of effort to triage or close issues. Providing a 
simple automated tool to detect potential unrelated questions can help project managers 
and software developers to focus on more practical issues.

Other works of software defect classification focus on classifying bugs, 
features, enhancements, etc. Most of them don't even include questions. Since most of them are 
based on the data provided from internal issue trackers of projects like JBoss, Apache Foundation, they are
trained on a structured and clean dataset. However, for developers that use the built-in issue tracker
of platforms like GitHub, the problem still remains. Moreover, we want to focus on implementing 
a binary classifier to filter questions, instead of the multi-class classifiers that other 
works propose. Maybe we can say that the goals are a little different or even complementary. 
For example, we can first filter spam by using our proposed classifier then when 
we are more confident that the reported issue is related, we can use another classifier to 
automatically categorize it. This may improve the accuracy of automatic categorization too. Also, this 
enables deployment of those classifiers on a broader range of projects.

In this report, we want to investigate the feasibility of labeling unrelated questions
in the issue tracker of popular open source software on platforms like GitHub. We want to evaluate the performance
of different classification algorithms. And finally, to compare the performance of a binary classification 
(question, not-question) approach to other multi-class methods used in other works.

\begin{figure}[t]
  \centerline{\includegraphics[width=3.5in]{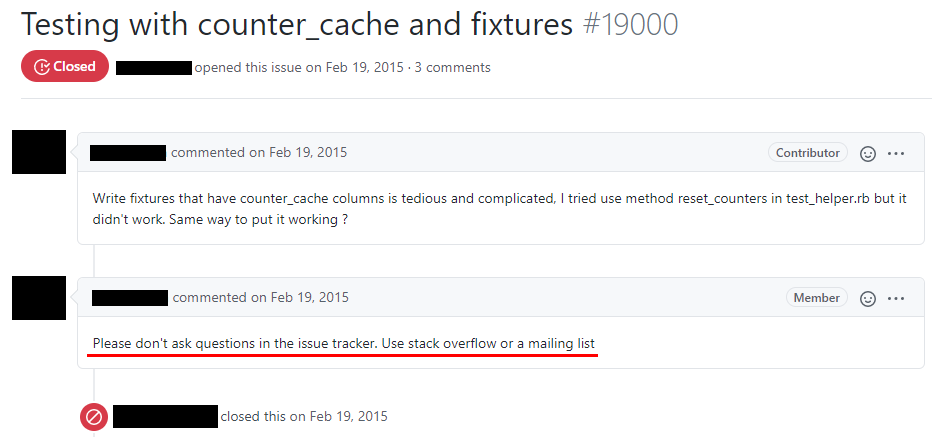}}
  \caption{A question submitted to the issue tracker of rails project on GitHub.}
  \label{fig:qissue}
\end{figure}

\begin{figure*}[t]
  \centerline{\includegraphics{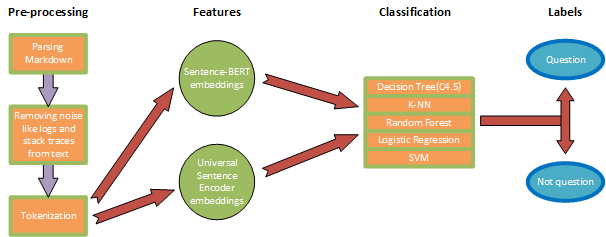}}
  \caption{Summary of our approach at a high level.}
  \label{fig:process}
\end{figure*}

As mentioned earlier, our goal is to train a binary classifier to automatically label questions.
We have used a previously available dataset of GitHub issues provided by \cite{8816794}.
Cleaning and pre-processing the dataset narrowed down the count of our available 
records to 102,198. The pre-processing part was the hardest part of our implementation
since the text of issues usually contain machine-generated texts like logs and stack traces. 
After extracting the text part of issues, we have used state-of-the-art sentence embedding techniques
to convert text data to numerical vectors. By evaluating several classification 
algorithms, we achieved the best result with the Logistic Regression algorithm.

First, In Section \ref{relworks} we discuss other related works. 
In Section \ref{approach} we describe our dataset, data pre-processing methods, 
and technical implementation. In Section \ref{eval} we show the performance of our implementation
and discuss the results of different classification methods.
Finally, Section \ref{conclusion} concludes this report.

\section{Related Works}\label{relworks}

The work of Antoniol et al. \cite{10.1145/1463788.1463819} is one of the defect classification works. They trained  
classifiers with Naive Bayes, ADTree, and Logistic Regression on the text data of 1,800 issues extracted 
from Mozilla, Eclipse, and JBoss. Their work is focused on classifying whether a submitted issue is bug or 
non-bug (enhancement, feature request, etc). Zhou et al. \cite{6976097} improve on   \cite{10.1145/1463788.1463819} 
by including other structured attributes available in an issue tracker,
such as assignee, dates, severity, into the machine learning model. Pingclasai et al. \cite{6754344} build a 
classifier using LDA topic modeling for reclassification of issues. 
They used Decision Tree, Naive Bayes classifier, and Logistic Regression to classify issues to bug and not-bug. 
Another work by Cubranic and Murphy \cite{1814517} use Naive Bayes for text classification 
to find an assignee for a bug report. Hanmin and Xin used an LSTM neural network \cite{10.1145/3275219.3275239}
to classify issues to bug and non-bugs too, they showed that with an LSTM neural network they can achieve better 
performance than some other machine learning based approaches.

Kochhar et al. \cite{6923127} address the problem of misclassified reports in issue trackers. They have 
extracted features from 7,000 issues from five open source projects to classify them into 13 categories. Unlike
the other mentioned works, Kochhar et al. build a multi-class classifier using textual and structural data from 
issues. But their categories do not include question.

To the best of our knowledge, we couldn't find any issue classification works which included questions. Furthermore, 
almost all of the mentioned works were done on a small dataset (less than 8,000 records) collected from internal 
issue trackers.

\section{Approach}\label{approach}
In summary, our approach involves three main phases:
\begin{enumerate}
  \item The first phase is cleaning and pre-processing the text data of issues. 
  We used around 400 regular expressions to remove noise from the text of issues. 
  Most of the issues are not properly formatted and they contain stack traces, log lines, code snippets, 
  command lines, environment variables, configurations, IP addresses, identifiers, flags, timestamps, etc.
  We have also applied the standard NLP pre-processing tasks such as tokenization and removing punctuations.
  \item The second phase is computing a sentence embedding of each text document provided by 
  the previous step. We have used two state-of-the-art sentence embedding algorithms to convert our
  documents to a high dimensional vector representation. These embedding algorithms have shown a high 
  level of performance for several tasks \cite{reimers-2019-sentence-bert}.
  \item The final phase is training a classifier on the extracted features. We have evaluated
  the performance of SVM, Decision Tree (C4.5), Logistic Regression, Random Forest, and k-NN 
  classification algorithms.
\end{enumerate}

In the following subsections, we explain our dataset, pre-processing methods, a few background definitions, and
our implementation.

\subsection{Dataset}
We used the RapidRelease \cite{8816794} dataset. It contains 2 million issues from active and popular 
GitHub repositories of 18 programming languages. The dataset is a SQL database.
The labels of each issue, which were applied by the project maintainers, are also available in this dataset.

With a simple query, we extracted frequently applied labels. You 
can see the count of the top 10 labels in Table \ref{tab:labels}. 
We have selected these top labels: \verb|bug|, \verb|duplicate|, \verb|enhancement|, \verb|wontfix|,
\verb|feature|, \verb|improvement|, and \verb|question| to find more labels based on them.
For example, some projects may use the label "type: bug" to represent a bug. Also, many of the issues 
have more than one label. The labels are stored in the database as a string in a comma-separated format. 
For example, an issue with both labels of bug and duplicate is stored as ``bug,duplicate". 
For each of these base labels, we queried issues that contained their strings. For example,
for features, we looked for labels that contained the string ``feature". We also queried their usage count and
extracted labels for each category based on their usage count.
For example, The top five applied labels that contained the ``feature" string 
are: ``feature", ``feature request", ``kind/feature", ``cat:feature", ``type: feature".
A label with a higher usage indicates that it is a relevant label for the intended purpose. This query resulted
in thousands of records for each of the base labels, however, many of them were only applied
a handful of times. Therefore we have only selected labels from these lists that 
at least were used 50 times. For our classifier, we have two categories of \verb|not-question| 
and \verb|question|. We used the base labels \verb|bug|, \verb|duplicate|, \verb|enhancement|, \verb|wontfix|,
\verb|feature|, \verb|improvement|, and their similar labels for the \verb|not-question| category. We used
the \verb|question| or similar labels for the \verb|question| category. Given that many issues have 
more than one label, we removed issues that had labels like ``bug,question'' or ``support,bug'' from 
the list of frequent labels. We used this list of labels to extract issues. The list of labels
used can be found in our code repository in the \verb|labels| folder for each one of the base labels.
With this process we extracted 46,549 issues for the \verb|question| category and 280,829 for 
the \verb|not-question|.

\begin{table}[t]
  \caption{Top 10 Applied Labels}
  \begin{center}
  \begin{tabular}{|c|c|}
  % \hline
  % \multicolumn{2}{|c|}{\textbf{Top 10 Applied Labels}} \\
  \hline
  % \cline{2-4} 
  \textbf{Label} & \textbf{Count}\\
  \hline
  bug           & 65735 \\
  \hline
  enhancement   & 43673 \\
  \hline
  question      & 34248 \\
  \hline
  cla: yes      & 17476 \\
  \hline
  duplicate     & 12210 \\
  \hline
  feature       & 9235 \\
  \hline
  stale         & 7996 \\
  \hline
  documentation & 7814 \\
  \hline
  invalid       & 7658 \\
  \hline
  cla signed    & 7306 \\
  \hline
  % \multicolumn{4}{l}{$^{\mathrm{a}}$Sample of a Table footnote.}
  \end{tabular}
  \end{center}
  \label{tab:labels}
\end{table}

\subsection{Date cleaning and pre-processing}
After extracting issues the pre-processing phase begins. GitHub issues are written in the Markdown format.
We used the \verb|marked| \cite{web:marked} library to parse and compile issues, this library compiles
Markdown to HTML. Following compiling issues to HTML we extracted the text part of issues from HTML.
Also, during the translation to HTML, we removed image, table, code, pre, and header tags. If the
author of an issue had used the Markdown properly then after removing these tags only the text part
of an issue would have remained. However, many developers do not submit issues in the proper format and we
still had to further clean texts. 

Next we concatenated the title and body of issues then applied more than 400 regular expressions 
to remove the remaining noise from the text of issues. Specifically,
we have removed dependency trees, emojis, code snippets, comments, logs,
error message, stack traces, timestamp, date times, command lines, environment variables, identifiers,
HTML, markup tags, module versions, IP addresses, emails, GitHub user handles, URIs, file paths, and digits.
These regular expressions were designed by manually skimming through texts. Suppose we want to extract
stack traces, most of the stack traces contain several consecutive lines that begin with 
the ``at'' or ``in'' word or they contain the file extension of that specific 
programming language (``.js'', ``.py''). Also, stack traces of a platform like Java or .NET are 
very similar to each other and creating a pattern to extract them is easy.
By manually analyzing our regular expression for stack traces we identified thousands
of occurrences. As another example, suppose we want to remove log lines. Many developers copy logs or debug
output of the software to help diagnose the problem. Almost all of the logs contain a timestamp on
each line, therefore, if we see several consecutive lines that contain timestamps, there is 
a high probability that these lines are software generated log outputs.

By manually applying our regular expressions and inspecting dozens of results we have fine tuned 
them to not catch incorrect patterns. Of course, these regular expressions are not perfect and there are 
still a handful of incorrect matches in a large dataset, but given that they can match hundreds of correct patterns, 
we opted to apply them to remove noise from the text of issues. We have also removed several punctuations except 
those that indicated the end of a sentence (such as .,;!?). The list of all used regular expressions 
and the script is available in the code repository.

The next step of pre-processing was filtering issues that were not in the English language. We used 
the \verb|langdetect| \cite{web:langdetect} library to determine the language of each issue 
and then filter the ones that were not English.

Finally, we used the nltk library for tokenization. We used its default tokenizer
which is \verb|TreebankWordTokenizer|. Following 
tokenization, we marked issues that had more than five and lower than 200 tokens for training. 
This reduced the records of the \verb|question| category 
to 42,198 and the \verb|not-question| category to 262,425. Also, for the \verb|question| category we only selected
issues that were closed.

\subsection{Sentence Embeddings}
To understand the concept of sentence embeddings we should first understand word embeddings.
Word embeddings are essentially vector representations of words. There are a few techniques to train
numerical vector representation of words. Word2vec \cite{mikolov2013efficient,mikolov2013distributed} 
and GloVe \cite{pennington2014glove} are the most famous word embeddings algorithms. The advantage of
word embeddings over older techniques like One-hot encoding and Bag-of-words is that semantically similar
words appear close to each other in the vector space and the relation among the words are preserved.
One popular example of this case is the similarity between the words ``king'' and ``queen'', for example,
we can calculate a vector close to the vector of word ``queen'' by calculating the formula:
\begin{equation}
  king-men+woman=queen
\end{equation}
Word embeddings give us a numerical vector for each word. Sentence embeddings are an extension to 
the word embeddings technique. Sentence embeddings are vector representations of a sentence, paragraph, 
or even a whole text document. One naive way of calculating the sentence embeddings of a text or sentence 
is to average its word embeddings, however, this method does not perform well for NLP tasks. 
Sentence-BERT \cite{reimers-2019-sentence-bert} is a state-of-the-art 
algorithm to derive sentence embeddings. Sentence-BERT builds on top of the BERT \cite{devlin2019bert} networks 
to calculate meaningful sentence embeddings. Another algorithm to calculate sentence embeddings is 
Universal Sentence Encoder \cite{cer2018universal}. Vectors generated by both of the two mentioned sentence
embedding algorithms can be compared using the cosine similarity function. Sentence embeddings perform
extremely well for similarity and classification tasks. Since our task is classification, we decided to 
use these sentence embeddings to compute a numerical vector for the text of each issue in our 
cleaned dataset. We computed embeddings using both Universal Sentence Encoder and Sentence-BERT then we trained
classifiers for both of them to compare results. The input to these models is a string and the output is 
a numerical vector. We used the \verb|roberta-large-nli-stsb-mean-tokens| pre-trained 
model for Sentence-BERT which is trained over the SNLI \cite{snli:emnlp2015} and Multi NLI \cite{N18-1101} datasets. 
This model generates 1024 dimension vectors. We used the 
\verb|universal-sentence-encoder-large-v5| pre-trained model of Universal Sentence Encoder algorithm. This model generates 
512 dimension vectors. This model is trained over a variety of unsupervised data sources from the web. They used data
from Wikipedia and question-answer websites. The resulting labeled vectors of both of these embeddings 
are provided in a \textit{csv} file.

\subsection{Training classifiers}
After extracting feature vectors we now have a dataset ready to train classification models. We have two datasets,
one generated from Universal Sentence Encoder and one generated from Sentence-BERT. The Universal Sentence Encoder
records have 512 dimensions while the Sentence-BERT ones have 1024 dimensions. We used the Weka UI toolkit 
\cite{10.1145/1656274.1656278} to train classifiers. We have trained classifiers using Logistic Regression, 
Decision Tree (C4.5), Random Forest, Support Vector Machine, and K-NN. In this section, we briefly review each of the 
classification algorithms.
\begin{itemize}
  \item\textbf{Logistic Regression:}: Logistic Regression can be used for binary classification. Instead of 
  predicting a continuous value like Linear Regression, it predicts the probability belonging to two different 
  classes. During the training process, it finds a logistic function to calculate this probability.
  \item\textbf{Decision Tree (C4.5):} The C4.5 decision tree algorithm calculates the information gain ratio 
  for all of the attributes then splits data at each node by the attribute with the highest normalized 
  information gain ratio. It recursively continues to split data until the tree is complete.
  \item\textbf{Random Forest:} Random Forest is an Ensemble Learning algorithm. Basically, it consists of several 
  uncorrelated decision trees, each used for classification. The class with the majority of votes becomes 
  the final results.
  \item\textbf{Support Vector Machine:} SVM is one of the best performing linear classification algorithms. It can 
  classify data by finding the best hyperplane that separates the dataset into two given classes. To allow 
  room for generalization it also selects the hyperplane which has the most margin from both categories. It can 
  also work on a non-linearly separable dataset by using a kernel function.
  \item\textbf{k-Nearest Neighbors:} To use the k-NN algorithm for classification the raw dataset becomes the model,
  there is no training process. When a record is queried for its class, the algorithm finds the \textit{k} closest 
  data points in the raw dataset according to a distance function (e.g. Euclidean distance). The class of 
  the queried record is determined by the majority vote of its neighbors. The number of neighbor 
  points (\textit{k}) to use for classification is an input parameter, the best \textit{k} value can be 
  determined by evaluating results over the training split.
\end{itemize}

\section{Empirical Evaluation}\label{eval}
We have implemented our approach to answer our three research questions. Our questions are:
\begin{itemize}
  \item{RQ1)} \textit{Is it possible to prevent spamming in issue trackers of popular open source software? 
  is it possible to label questions asked in issue trackers?}
  \item{RQ2)}\textit{Which classification algorithms yield goods result for this specific task?}
  \item{RQ3)} \textit{Is the binary classification (question, not-question) a good approach? 
  does it help or improve results?}
\end{itemize}

\subsection{RQ1 and RQ2}
In order to answer our first two questions, we have evaluated the performance of several popular
classification algorithms over both datasets. One dataset containing feature vectors generated by 
the Sentence-BERT algorithm and another generated by the Universal Sentence Encoder algorithm. We have two labels for
classification. The \verb|question| category has 42,198 issues. For the \verb|not-question| category, 
we randomly chose 60,000 issues from a total of 262,425. This made our training dataset more balanced
while not breaking the 60-40 threshold. In total, our dataset contains 102,198 records. The exact same issues were used 
for both of the sentence embedding algorithms, however, the generated feature vectors have different dimensions, 
vectors of the Sentence-BERT algorithm has 1024 dimension while vectors of the Universal Sentence Encoder has 
512 dimensions. Vectors generated by both these algorithms are normalized and we did not perform any normalization
on them for any of the classification algorithms. We used the Weka toolkit to train classification models. 
If Weka toolkit normalized values by default for any of the classification algorithms, 
we turned it off before training models. 

We trained classifiers with the Logistic Regression, Decision Tree (C4.5), Random Forest, and Support Vector Machine
algorithms. For all of the classifiers, we ran the training process with their default parameters in Weka. You can find 
these parameters in the code repository under the \verb|weka| folder. We trained all of the algorithms over 70\%
of the dataset and we evaluated them over the 30\% test split. Also, we trained each of these algorithms two 
times since we had two datasets. Due to the limited computation resources and time, we were not able to 
perform cross-validation. Still, a 30\% test split over a dataset with more than 100,000 records is large 
and diverse enough to give reliable results.
The output of Weka for each algorithm is available in our code repository too. 

In Table \ref{tab:results} you can find the accuracy of each classifier. Logistic Regression has the 
highest accuracy among Universal Sentence Encoder data and SVM has the highest accuracy among 
Sentence-BERT. The best classifier is the Logistic Regression trained over Universal Sentence Encoder 
embeddings with an accuracy of 81.68\%. It almost performs two percent better than the best classifier 
trained over the Sentence-BERT embeddings. In addition, we can observe that classifiers trained over 
the Universal Sentence Encoder embeddings have more consistent results. Almost all of its classifiers  
perform better with an accuracy of at least 78\% (except the decision tree).

\begin{table}[t]
  \caption{Accuracy of classification algorithms trained over both embeddings with a 70-30 split.}
  \begin{center}
  \begin{tabular}{|l|c|c|}
  % \hline
  % \multicolumn{2}{|c|}{\textbf{Top 10 Applied Labels}} \\
  \hline
  % \cline{2-4} 
  \textbf{Sentence Embedding} & \textbf{Classifier}&\textbf{Accuracy}\\
  \hline
                            &k-NN                & 78.38\% \\
                            &Decision Tree(C4.5) & 68.12\% \\
  Universal Sentence Encoder&\textbf{Logistic Regression} & \textbf{81.68\%} \\
  (512 features)            &Random Forest       & 79.05\% \\
                            &SVM                 & 78.18\% \\
  \hline
               &k-NN                & 69.88\% \\
               &Decision Tree(C4.5) & 60.41\% \\
  Sentence-BERT&Logistic Regression & 78.42\% \\
  (1024 features)&Random Forest     & 71.35\% \\
               &\textbf{SVM}        & \textbf{79.86\%}\\
\hline
  % \multicolumn{4}{l}{$^{\mathrm{a}}$Sample of a Table footnote.}
  \end{tabular}
  \end{center}
  \label{tab:results}
\end{table}

\begin{figure*}[t]
  \centerline{\includegraphics[width=7.16in]{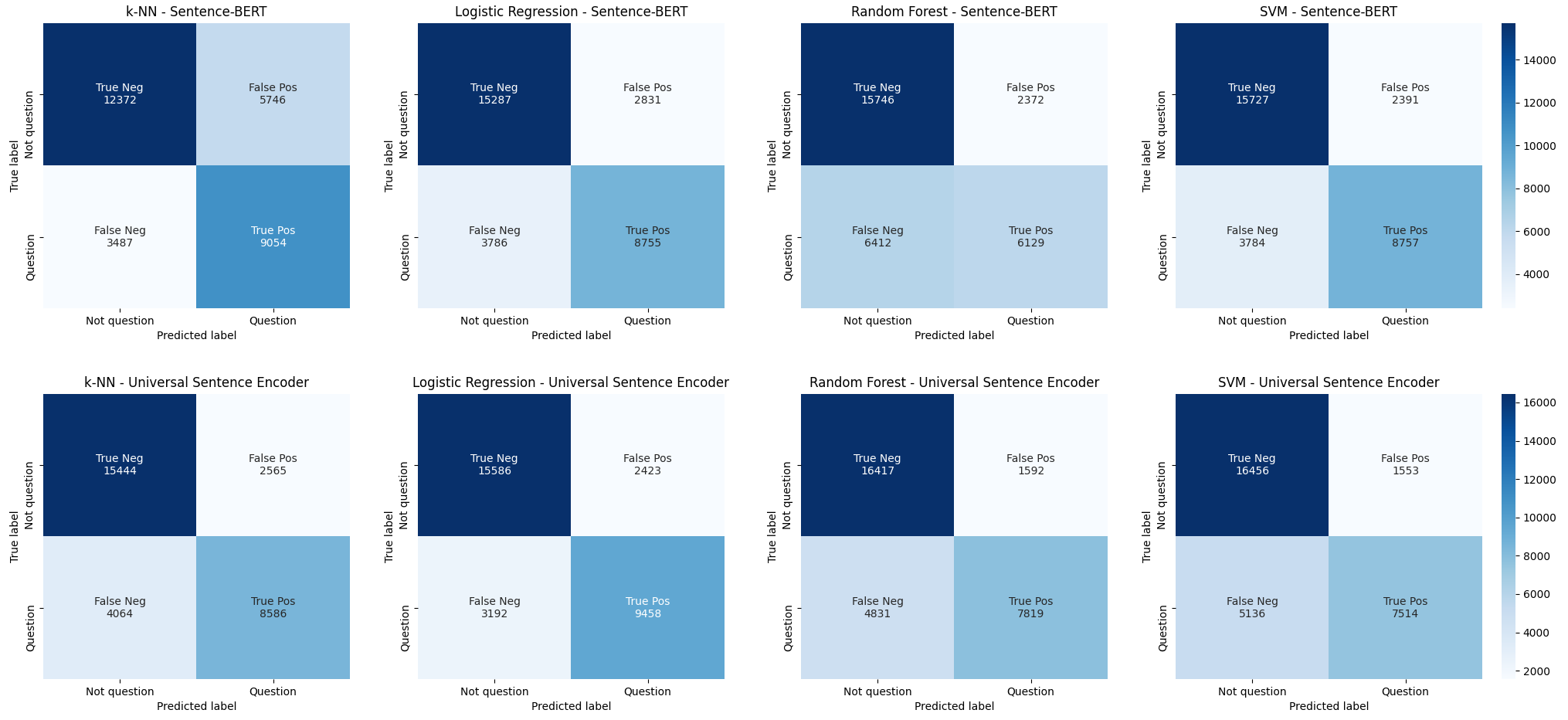}}
  \caption{Confusion matrix of trained models over both sentence embedding techniques.}
  \label{fig:cfheatmap}
\end{figure*}

\begin{table}[h!b]
  \caption{Precision and recall rates of }
  \begin{center}
  \begin{tabular}{|l|c|c|c|c|}
  \hline
  \textbf{Classifier}&\textbf{Class}&\textbf{Precision}&\textbf{Recall}&\textbf{F-Measure}\\
  \hline
  \multicolumn{5}{|c|}{\textit{Universal Sentence Encoder}}\\
  \hline
  k-NN               &NQ&79.2\%&85.8\%&82.3\%\\
                     &Q &77.0\%&67.9\%&72.1\%\\
  \hline
  Decision Tree(C4.5)&NQ&73.0\%&72.6\%&72.8\%\\
                     &Q &61.3\%&61.8\%&61.5\%\\
  \hline
  Logistic Regression&NQ&83.0\%&86.5\%&84.7\%\\
                     &Q &79.6\%&74.8\%&77.1\%\\
  \hline
  Random Forest      &NQ&77.3\%&91.2\%&83.6\%\\
                     &Q &83.1\%&61.8\%&70.9\%\\
  \hline
  SVM                &NQ&76.2\%&91.4\%&83.1\%\\
                     &Q &82.9\%&59.4\%&69.2\%\\
  \hline
  \multicolumn{5}{|c|}{\textit{Sentence-BERT}}\\
  \hline
  k-NN               &NQ&78.0\%&68.3\%&72.8\%\\
                     &Q &61.2\%&72.2\%&66.2\%\\
  \hline                     
  Decision Tree(C4.5)&NQ&66.7\%&66.0\%&66.3\%\\
                     &Q &51.6\%&52.3\%&51.9\%\\
  \hline                   
  Logistic Regression&NQ&80.1\%&84.4\%&82.2\%\\
                     &Q &75.6\%&69.8\%&72.6\%\\ 
  \hline
  Random Forest      &NQ&71.1\%&86.9\%&78.2\%\\ 
                     &Q &72.1\%&48.9\%&58.3\%\\
  \hline                   
  SVM                &NQ&80.6\%&86.8\%&83.6\%\\
                     &Q &78.6\%&69.8\%&73.9\%\\ 
  \hline
  \end{tabular}
  \end{center}
  \label{tab:results2}
\end{table}
To evaluate our approach better we examine other metrics too, accuracy alone is not a good metric for 
evaluation. We have gathered the confusion matrix of each classifier in Fig.~\ref{fig:cfheatmap}. 
By looking at it we see that the hardest part is labeling actual questions, all of the algorithms almost 
struggle at classifying them. Logistic Regression over Universal Sentence Encoder embeddings, which had 
the highest accuracy, has the highest TP rate for the question category too (74.8\%). SVM over 
Universal Sentence Encoder embeddings has the highest true negatives category. 
But The SVM classifier also has the lowest false positive numbers, the Logistic Regression has almost 1000 more 
false positives. In classifiers trained over the Sentence-BERT embeddings, SVM has both the highest TP (69.8\%) 
and the lowest FP rate (13.2\%). But the Random Forest classifier has the highest true negatives with a 
little margin higher than the SVM.

To further compare the results, we have provided Precision, Recall, and F-measure rates in Table \ref{tab:results2}.
The Logistic algorithm trained over Universal Sentence Encoder embeddings has the highest F-measure rates.
In the Sentence-BERT embeddings category, the SVM classifier has the highest F-measure. By analyzing the precision 
rates we can still see that Logistic Regression trained with universal Sentence Encoder embeddings still has 
the best performance, its average precision is higher than any other classifier. 
Although, as we can see in Fig.~\ref{fig:cfheatmap} it has much higher false positives than the SVM and 
Random Forest. It may be more sensible to use these two instead of the Logistic Classifier.

Regarding RQ1, given the evaluations and empirical evidence, we can conclude that it is feasible to classify 
questions in issue trackers of platforms like GitHub with pretty good precision. We saw that we can achieve 
an accuracy rate of 81.6\%. To answer RQ2 we trained several models with popular algorithms. We assessed two
different Sentence Embedding techniques to find which is more suitable for our task. We observed that the 
Universal Sentence Encoder has better and more consistent performance for our task. The answer to RQ2 is not a 
single algorithm, if we only measure the accuracy metric, the Logistic Regression works best. However, if we
analyze other metrics we may choose other algorithms. For example, the Random Forest model trained with
Universal Sentence Encoder embeddings has an accuracy of 79.00\% but it has much lower false positives than the 
Logistic Regression classifier.

We noticed that the bottleneck in the accuracy of almost all of the classifiers is categorizing issues that are 
actually questions. They can label issues that are not questions with good accuracy. This problem can be attributed 
to the fact that many of the submitted questions in issue trackers will be converted to another type of issue by 
the maintainers. For example, a submitted question about a feature may interest developers and they decide to 
implement it, thus they will remove the ``question'' label and apply a ``feature'' label to the issue which is 
asking a question. The same happens with other labels such as ``bug'' too, a submitted question about an error 
may in fact turn out to be a bug. Since our dataset is not manually labeled for the purpose of our task and we 
depend on the labels provided by authors of projects this problem can't be addressed trivially. 
Manually labeling issues may help address this problem but it is time consuming. Another approach may be to analyze 
issues with an unsupervised algorithm.

\subsection{RQ3}
With RQ3 we are interested in comparing our approach with other works. However, we found out 
a direct comparison with other works is not possible. As we saw in Section \ref{relworks}, almost 
all of the works in the defect classification area are multi-class classification problems. 
Nevertheless, we think that our approach and other works are complementary.
We can apply both of them in a pipeline fashion. First, we can separately label questions submitted to an 
issue tracker, then apply another categorization method to predict the class of defect.

\subsection{Limitations and Further Improvements}
\label{limits}
Due to the size of our dataset and limited computation power, we were not able to tune hyperparameters of 
classification algorithms used in Weka. We used the default parameters set by Weka. 
However, when we searched for the best value for a few of the parameters over a small sample dataset,
we saw a modest improvement in results.

Our initial solution to solve this problem was to train an LSTM network with Word Embeddings. Using lower level 
embeddings with a neural network may improve results. But we don't know how much improvement it can yield.
As explained in Section \ref{approach} after pre-processing the RapidRelease dataset, we only used issues with 
more than five and less than 200 tokens to train models. Also, vectors generated by Word Embedding algorithms 
can have at least 100 dimensions. With this high amount of input and training parameters, training a neural network 
failed on our device. 

As mention above, we filtered issues with long texts, we have only trained models over issues that had lower than 200 
tokens. We removed longer issues because the performance of sentence embedding algorithms degrades 
with longer texts. Still, 200 tokens are long enough for most of the issues and we also counted tokens after 
removing all the noise (log lines, stack traces) from the text of issue.

\subsection{Threats to Validity}
Like all evaluations, our tests are also prone to external validity and generalization. We have only used data from 
the GitHub issue tracker. But we think that the number of questions asked on the issue tracker of projects that don't 
use GitHub (for example, ASF software have their own issue trackers) is very low. 
Furthermore, the RapidRelease dataset is diverse and includes issues from several developer communities 
and programming languages. 

Another factor is that we relied on the labels applied by developers of projects. The labels may not completely 
reflect our intended two categories, especially for questions. As we discussed in Subsection \ref{limits} this 
may be the reason why classifiers had less precision for the question category. Overall, we tried to use labels 
that were frequently applied by developers to issues.

\section{Lessons Learned}
The most important lesson I learned was to not rely on a single solution. It is always best to evaluate several 
approaches to find the best possible fit. Another thing I noticed is that the accuracy metric alone is not enough 
to gauge performance. We have to look at other metrics too and analyze more details. The accuracy metric can be 
misleading.

\section{Conclusion}\label{conclusion}
Our goal was analyzing feasibility of detecting unrelated questions submitted to issue trackers in platforms like 
GitHub. To address this problem We used a previously provided dataset of GitHub issues by the RapidRelease work. 
We filtered and pre-processed the raw dataset and narrowed count of issues for analysis to approximately 
102,000 issues. We were able to filter noise, such as stack traces, logs, etc, from text of issues. 
To prepare our textual data for machine learning algorithms, we used two of state-of-the-art sentence embeddings
techniques, Sentence-BERT and Universal Sentence Encoder, to extract numerical feature vectors. Finally, we 
trained classifiers over our dataset with five famous classification algorithms. The best result was yielded
with the Logistic Regression algorithm trained over Universal Sentence Encoder embeddings with the accuracy
rate of 81.68\%. Overall, we conclude that it is possible to classify questions submitted to a public issue tracker
with a good probability. 

\bibliographystyle{IEEEtran}
\bibliography{IEEEabrv,references}

\begin{thebibliography}{10}
\providecommand{\url}[1]{#1}
\csname url@samestyle\endcsname
\providecommand{\newblock}{\relax}
\providecommand{\bibinfo}[2]{#2}
\providecommand{\BIBentrySTDinterwordspacing}{\spaceskip=0pt\relax}
\providecommand{\BIBentryALTinterwordstretchfactor}{4}
\providecommand{\BIBentryALTinterwordspacing}{\spaceskip=\fontdimen2\font plus
\BIBentryALTinterwordstretchfactor\fontdimen3\font minus \fontdimen4\font\relax}
\providecommand{\BIBforeignlanguage}[2]{{%
\expandafter\ifx\csname l@#1\endcsname\relax
\typeout{** WARNING: IEEEtran.bst: No hyphenation pattern has been}%
\typeout{** loaded for the language `#1'. Using the pattern for}%
\typeout{** the default language instead.}%
\else
\language=\csname l@#1\endcsname
\fi
#2}}
\providecommand{\BIBdecl}{\relax}
\BIBdecl

\bibitem{8816794}
S.~D. {Joshi} and S.~{Chimalakonda}, ``Rapidrelease - a dataset of projects and issues on github with rapid releases,'' in \emph{2019 IEEE/ACM 16th International Conference on Mining Software Repositories (MSR)}, 2019, pp. 587--591.

\bibitem{10.1145/1463788.1463819}
\BIBentryALTinterwordspacing
G.~Antoniol, K.~Ayari, M.~Di~Penta, F.~Khomh, and Y.-G. Gu\'{e}h\'{e}neuc, ``Is it a bug or an enhancement? a text-based approach to classify change requests,'' in \emph{Proceedings of the 2008 Conference of the Center for Advanced Studies on Collaborative Research: Meeting of Minds}, ser. CASCON '08.\hskip 1em plus 0.5em minus 0.4em\relax New York, NY, USA: Association for Computing Machinery, 2008. [Online]. Available: \url{https://doi.org/10.1145/1463788.1463819}
\BIBentrySTDinterwordspacing

\bibitem{6976097}
Y.~{Zhou}, Y.~{Tong}, R.~{Gu}, and H.~{Gall}, ``Combining text mining and data mining for bug report classification,'' in \emph{2014 IEEE International Conference on Software Maintenance and Evolution}, 2014, pp. 311--320.

\bibitem{6754344}
N.~{Pingclasai}, H.~{Hata}, and K.~{Matsumoto}, ``Classifying bug reports to bugs and other requests using topic modeling,'' in \emph{2013 20th Asia-Pacific Software Engineering Conference (APSEC)}, vol.~2, 2013, pp. 13--18.

\bibitem{1814517}
D.~Cubranic and G.~C. Murphy, ``Automatic bug triage using text categorization,'' in \emph{Software Engineering and Knowledge Engineering}, 2004, p. 92{\textendash}97.

\bibitem{10.1145/3275219.3275239}
\BIBentryALTinterwordspacing
H.~Qin and X.~Sun, ``Classifying bug reports into bugs and non-bugs using lstm,'' in \emph{Proceedings of the Tenth Asia-Pacific Symposium on Internetware}, ser. Internetware '18.\hskip 1em plus 0.5em minus 0.4em\relax New York, NY, USA: Association for Computing Machinery, 2018. [Online]. Available: \url{https://doi-org.proxy.bib.uottawa.ca/10.1145/3275219.3275239}
\BIBentrySTDinterwordspacing

\bibitem{6923127}
P.~S. {Kochhar}, F.~{Thung}, and D.~{Lo}, ``Automatic fine-grained issue report reclassification,'' in \emph{2014 19th International Conference on Engineering of Complex Computer Systems}, 2014, pp. 126--135.

\bibitem{reimers-2019-sentence-bert}
\BIBentryALTinterwordspacing
N.~Reimers and I.~Gurevych, ``Sentence-bert: Sentence embeddings using siamese bert-networks,'' in \emph{Proceedings of the 2019 Conference on Empirical Methods in Natural Language Processing}.\hskip 1em plus 0.5em minus 0.4em\relax Association for Computational Linguistics, 11 2019. [Online]. Available: \url{https://arxiv.org/abs/1908.10084}
\BIBentrySTDinterwordspacing

\bibitem{web:marked}
\BIBentryALTinterwordspacing
C.~Jeffrey, ``Marked,'' 2020. [Online]. Available: \url{https://marked.js.org/}
\BIBentrySTDinterwordspacing

\bibitem{web:langdetect}
\BIBentryALTinterwordspacing
M.~Danilak, ``langdetect,'' 2020. [Online]. Available: \url{https://github.com/Mimino666/langdetect/}
\BIBentrySTDinterwordspacing

\bibitem{mikolov2013efficient}
T.~Mikolov, K.~Chen, G.~Corrado, and J.~Dean, ``Efficient estimation of word representations in vector space,'' 2013.

\bibitem{mikolov2013distributed}
T.~Mikolov, I.~Sutskever, K.~Chen, G.~Corrado, and J.~Dean, ``Distributed representations of words and phrases and their compositionality,'' 2013.

\bibitem{pennington2014glove}
\BIBentryALTinterwordspacing
J.~Pennington, R.~Socher, and C.~D. Manning, ``Glove: Global vectors for word representation,'' in \emph{Empirical Methods in Natural Language Processing (EMNLP)}, 2014, pp. 1532--1543. [Online]. Available: \url{http://www.aclweb.org/anthology/D14-1162}
\BIBentrySTDinterwordspacing

\bibitem{devlin2019bert}
J.~Devlin, M.-W. Chang, K.~Lee, and K.~Toutanova, ``Bert: Pre-training of deep bidirectional transformers for language understanding,'' 2019.

\bibitem{cer2018universal}
D.~Cer, Y.~Yang, S.~yi~Kong, N.~Hua, N.~Limtiaco, R.~S. John, N.~Constant, M.~Guajardo-Cespedes, S.~Yuan, C.~Tar, Y.-H. Sung, B.~Strope, and R.~Kurzweil, ``Universal sentence encoder,'' 2018.

\bibitem{snli:emnlp2015}
S.~R. Bowman, G.~Angeli, C.~Potts, and C.~D. Manning, ``A large annotated corpus for learning natural language inference,'' in \emph{Proceedings of the 2015 Conference on Empirical Methods in Natural Language Processing (EMNLP)}.\hskip 1em plus 0.5em minus 0.4em\relax Association for Computational Linguistics, 2015.

\bibitem{N18-1101}
\BIBentryALTinterwordspacing
A.~Williams, N.~Nangia, and S.~Bowman, ``A broad-coverage challenge corpus for sentence understanding through inference,'' in \emph{Proceedings of the 2018 Conference of the North American Chapter of the Association for Computational Linguistics: Human Language Technologies, Volume 1 (Long Papers)}.\hskip 1em plus 0.5em minus 0.4em\relax Association for Computational Linguistics, 2018, pp. 1112--1122. [Online]. Available: \url{http://aclweb.org/anthology/N18-1101}
\BIBentrySTDinterwordspacing

\bibitem{10.1145/1656274.1656278}
\BIBentryALTinterwordspacing
M.~Hall, E.~Frank, G.~Holmes, B.~Pfahringer, P.~Reutemann, and I.~H. Witten, ``The weka data mining software: An update,'' \emph{SIGKDD Explor. Newsl.}, vol.~11, no.~1, p. 10–18, Nov. 2009. [Online]. Available: \url{https://doi.org/10.1145/1656274.1656278}
\BIBentrySTDinterwordspacing

\end{thebibliography}
\vspace{12pt}
\color{red}

\end{document}